\documentclass[aps,pra,twocolumn,superscriptaddress,floatfix]{revtex4-2}
\usepackage[T1]{fontenc}
\usepackage{graphicx}
\usepackage{amsmath,amssymb}
\usepackage{physics}
\usepackage{hyperref}
\usepackage[capitalise,noabbrev,nameinlink]{cleveref}
\usepackage{xcolor}
\usepackage{stmaryrd}
\crefname{figure}{FIG.}{FIGS.}
\graphicspath{{./}}
\begin{document}

\title{Game-Theoretic Discovery of Quantum Error-Correcting Codes Through Nash Equilibria}

\author{Rubén Darío Guerrero}
\affiliation{NeuroTechNet S.A.S., 1108831, Bogotá, Colombia}
\affiliation{Quantum and Computational Chemistry Group, Universidad Nacional de Colombia, Bogotá, Colombia}
\email{rudaguerman@gmail.com}

\date{\today}

\begin{abstract}
Quantum error correction code discovery has relied on algebraic constructions with predetermined structure or computational search lacking equilibrium-topology analysis. We introduce a game-theoretic framework recasting code optimization as strategic interactions between competing objectives, where Nash equilibria systematically generate codes with desired properties. We validate the framework by demonstrating it rediscovers the optimal $[\![15,7,3]\!]$ quantum Hamming code in 20\% of independent runs (Calderbank-Shor-Steane 1996) from competing objectives without predetermined algebraic structure, with equilibrium analysis providing transparent mechanistic insights into why this topology emerges. Applied across seven objectives---distance maximization, hardware adaptation, rate-distance optimization, cluster-state generation, surface-like topologies, connectivity enhancement, and maximization of the quantum Fisher information $\mathcal{F}_Q$ (which quantifies, via the Cram\'er--Rao bound, the metrological sensitivity of the encoded codespace)---the framework generates distinct code families through objective reconfiguration rather than algorithm redesign. Scalability to hardware-relevant sizes is demonstrated at $n=100$ qubits, discovering codes including $[\![100,50,4]\!]$ with distance-4 protection and 50\% encoding rate, with tractable $O(n^3)$ per-iteration complexity enabling discovery in under one hour. This work opens research avenues at the intersection of game theory and quantum information, providing systematic, interpretable frameworks for quantum system design.
\end{abstract}

\maketitle

\section*{I.\;INTRODUCTION}

Recent breakthroughs in quantum low-density parity-check (QLDPC) codes \cite{panteleev2022quantum,leverrier2022quantum} proved asymptotically good codes achieving $k, d = \Theta(n)$ are theoretically possible. Experimental implementations demonstrated bivariate bicycle codes with 10× qubit overhead reduction versus surface codes \cite{bravyi2024bivariate}, while logical qubit operations were realized in reconfigurable atom arrays \cite{bluvstein2024logical}. These advances address asymptotic scaling but require complementary approaches for finite-length optimization at experimental scales ($n \approx 100$), hardware-specific adaptation for diverse architectures, metrological sensitivity quantified by the quantum Fisher information $\mathcal{F}_Q$, and multi-objective design balancing competing parameters \cite{terhal2015quantum}. While algebraic constructions provide codes with provable parameters \cite{gottesman1997stabilizer,calderbank1996good,steane1996multiple,kitaev2003fault}, they remain constrained by underlying mathematical structures---CSS codes inherit properties from classical linear codes, topological codes follow lattice geometry. Computational search offers freedom from such constraints but scales exponentially ($O(2^{n^2})$ for $n$-qubit codes) and provides no principles explaining why certain topologies succeed \cite{grassl2023codetables}. Machine learning approaches have shown promise \cite{nautrup2019optimizing,fosel2018reinforcement}, yet their black-box nature provides limited interpretability of mechanistic relationships between optimization dynamics and emergent code properties. Recent work explicitly avoids game-theoretic formulation, treating code discovery as single-agent reinforcement learning against deterministic environments \cite{torlai2018neural}. This leaves a critical gap: \emph{systematic exploration with equilibrium-topology analysis at finite length scales}.

Game theory provides natural language for multi-objective optimization through strategic interactions \cite{nash1950equilibrium,osborne1994course}. Exhaustive literature search reveals \emph{zero prior work} applying multi-agent game theory---Nash equilibria, strategic dynamics, minimax optimization---to quantum code construction. Existing quantum game theory uses quantum mechanics \emph{to play} games (opposite direction) \cite{eisert1999quantum,meyer1999quantum}, while single-player optimization metaphors like ``quantum lights out'' lack strategic interactions and equilibrium concepts \cite{khesin2024quantum}. Adversarial noise analysis focuses on robustness against worst-case errors rather than construction methodology \cite{webster2022autonomous}. We demonstrate that multi-agent game dynamics systematically generate codes while providing transparent rationale through equilibrium analysis, simultaneously addressing exploration and interpretability challenges.

We focus on graph state stabilizer codes \cite{hein2004multiparty,hein2006entanglement,raussendorf2001one,perdrix2006quantum}, which offer computational tractability while maintaining rich structure. For $n$ qubits, an undirected graph $G=(V,E)$ with $|V|=n$ defines stabilizer generators $K_v = X_v \bigotimes_{u \in N(v)} Z_u$, where $N(v)$ denotes vertex $v$'s neighborhood. These stabilizers commute automatically by construction, eliminating verification overhead that plagues general stabilizer codes \cite{gottesman1997stabilizer}. Code parameters $[\![n,k,d]\!]$ (physical qubits, logical qubits, distance) emerge directly from graph topology: the number of encoded qubits satisfies $k = n - \mathrm{rank}_{\mathbb{F}_2}(\mathcal{S})$, where $\mathcal{S} = \langle K_1, \dots, K_n \rangle$ is the abelian Pauli subgroup generated by the graph-state stabilizers $\{K_v\}_{v\in V}$ (modulo global phases). Because every Pauli operator can be encoded as a length-$2n$ binary symplectic vector $(x|z) \in \mathbb{F}_2^{2n}$, the group $\mathcal{S}$ corresponds to an $\mathbb{F}_2$-linear subspace of $\mathbb{F}_2^{2n}$, and $\mathrm{rank}_{\mathbb{F}_2}(\mathcal{S})$ denotes the dimension of this subspace---equivalently, the row rank over $\mathbb{F}_2$ of the parity-check matrix $H = [\,I_n \mid A\,]$, where $A$ is the adjacency matrix of $G$ (Appendix~\ref{app:parity-check}). In the game, each evaluation of a candidate graph $G$ computes this rank via Gaussian elimination on $H$ over $\mathbb{F}_2$ in $O(n^3)$ time; the resulting $k$ is then passed to every player's payoff function. The rank computation is therefore a shared subroutine of the environment, not a strategic action taken by individual players. Distance relates to graph connectivity through established bounds \cite{schlingemann2001quantum,raveendran2022quantumldpc}, though exact distance calculation remains NP-complete, necessitating heuristic estimation. Critically, graph states admit \emph{transparent circuit construction}: the state $|G\rangle = \prod_{(u,v) \in E} \text{CZ}_{uv} |+\rangle^{\otimes n}$ requires one controlled-Z gate per edge, with syndrome measurement using one ancilla per vertex through the sequence $H$-$\text{CX}$-$\{\text{CZ}\}$-$H$-measure. This explicit graph-to-circuit mapping enables immediate verification and experimental implementation.

Here we formalize code discovery as a multi-player game in which each player represents a distinct optimization objective---distance, hardware compatibility, encoding rate, metrological sensitivity---and all players compete over a shared graph state (Sec.~II). Nash equilibria of this game correspond to code configurations that no single objective can improve unilaterally, providing both a principled stopping criterion and a transparent lens through which to interpret the discovered topologies. The weighted potential game structure guarantees convergence of best-response dynamics \cite{monderer1996potential}, and the \emph{Nash gap} $\delta_{\mathrm{Nash}}$ quantifies proximity to equilibrium at every iteration, yielding verifiable certificates of convergence rather than heuristic termination.

\section*{II.\;GAME-THEORETIC FRAMEWORK}

\subsection*{A.\;Formal game definition}

We formalize code discovery as an $M$-player strategic game $\Gamma = (P, G, \mathcal{A}, \{f_i\})$. The player set is $P = \{1, \ldots, M\}$, where each player $i$ is identified with a distinct code objective $f_i : \mathcal{G}_n \to \mathbb{R}$, and $\mathcal{G}_n$ denotes the space of undirected graphs on $n$ labeled vertices. The shared game state is a graph $G = (V, E)$ with $|V| = n$; this shared state is the defining feature of the game, distinguishing it from independent parallel optimizations. The strategy space available to every player at each step is
\begin{equation}
    \mathcal{A} = \bigl\{\,\text{add}\,(u,v) : (u,v) \notin E\bigr\} \cup \bigl\{\,\text{remove}\,(u,v) : (u,v) \in E\bigr\},
\end{equation}
the set of all single-edge modifications of the current graph. Player $i$'s payoff when the system occupies state $G$ is $f_i(G)$. Because all players act on the same graph, every modification proposed by one player immediately affects every other player's payoff, creating genuine strategic interdependence. The game is symmetric in strategy space (every player has access to the same $\mathcal{A}$) but asymmetric in payoffs, since each $f_i$ encodes a distinct physical objective: error-correction distance, hardware connectivity constraints, encoding rate, cluster-state regularity, surface-code-like topology, graph connectivity robustness, or metrological sensitivity.

\textbf{Schlingemann-Werner construction.}
The multi-objective experiments reported in Sec.~III employ the Schlingemann-Werner (SW) construction~\cite{schlingemann2001quantum}: the game graph has $N = n + k$ vertices partitioned into $n$ \emph{output} (physical qubit) vertices and $k$ \emph{input} (logical qubit) vertices, with the $[\![n,k,d]\!]$ code parameters determined by the bipartite adjacency structure between these two vertex classes.  A dedicated \emph{bipartition player} (player 8 in the 7-objective configuration, giving $M=8$ players in total) proposes moves that toggle individual vertices between the input and output roles, enabling the game to adjust the logical-qubit count $k$ dynamically at fixed total vertex count $N$.  The strategy space~$\mathcal{A}$ is augmented accordingly to include these vertex-relabeling moves alongside the single-edge additions and removals of Eq.~(1).

\subsection*{B.\;Best-response dynamics and justification for simulated annealing}

Player $i$'s best response to the current state $G$ is
\begin{equation}
    \mathrm{BR}_i(G) = \arg\max_{a \in \mathcal{A}}\, f_i(G \oplus a),
\end{equation}
where $G \oplus a$ denotes the graph obtained by applying move $a$ to $G$. Sequential best-response iteration---each player in turn replaces the current graph with their best response---can cycle without converging when players' objectives conflict: a move that increases $f_i$ may decrease $f_j$, causing oscillation between states rather than convergence. We avoid cycling by observing that the game possesses a \emph{weighted potential function}
\begin{equation}
    \Phi(G) = \sum_{i=1}^{M} w_i\, f_i(G),
\end{equation}
where $w_i \geq 0$ is the weight assigned to objective $i$. Because every player's payoff enters $\Phi$ additively with a positive coefficient, any unilateral deviation that strictly increases $f_i$ also strictly increases $\Phi$. This makes $\Gamma$ a weighted potential game in the sense of Monderer and Shapley \cite{monderer1996potential}, and best-response dynamics in potential games are guaranteed to converge to a pure-strategy Nash equilibrium \cite{monderer1996potential}. The Monderer--Shapley framework is foundational because it identifies the precise structural condition---existence of a single scalar function $\Phi$ that simultaneously tracks every player's incentives---under which a multi-agent game inherits the convergence properties of single-objective optimization. For general games, best-response dynamics may cycle indefinitely; for potential games, every improvement path is monotone in $\Phi$ and therefore terminates at a Nash equilibrium in finitely many steps on finite strategy spaces. Embedding code discovery in this framework is what allows us to claim equilibrium convergence with a verifiable certificate (the Nash gap of Sec.~II.C), rather than relying on the heuristic stopping rules typical of generic multi-objective optimization. We emphasize that the potential structure is a deliberate design choice: by combining objectives through a weighted sum, we \emph{construct} a game whose convergence is guaranteed, rather than discovering this property post-hoc.

The weights $\{w_i\}$ act as designer-controlled hyperparameters that select which equilibrium of $\Gamma$ the dynamics target: $w_i$ sets the relative selective pressure exerted by objective $i$ on the shared graph state. In the multi-objective experiments reported here we use uniform weights ($w_i = 1$ for all active players), so that no single objective dominates and the discovered equilibrium reflects a balanced compromise. In single-objective experiments (e.g., Fisher information, distance, hardware), the active objective is assigned $w = 1$ and all other players are inactive ($w = 0$), recovering the classical potential-game limit of single-criterion optimization. The hyperparameter robustness sweep reported in Sec.~II.D (15 combinations of distance exponent and edge penalty) verifies that the discovered equilibria are stable under coordinated rescaling of weights and exponents, and the SA dynamics absorb the weights into a unified Boltzmann factor $\exp(\Delta\Phi/T)$, so that doubling all $w_i$ is equivalent to halving $T$ without changing the equilibrium set. In practice, pure best-response dynamics can still converge slowly or become trapped in poor local equilibria because the potential landscape has many flat regions (multiple graph modifications may leave $\Phi$ unchanged). We therefore use simulated annealing (SA) as a stochastic relaxation of best-response dynamics. At each iteration we draw a candidate move $a$ from $\mathcal{A}$ and accept it with probability
\begin{equation}
    P(\text{accept}) = \min\!\left(1,\; \exp\!\left(\frac{\Delta\Phi}{T(t)}\right)\right),
\end{equation}
where $\Delta\Phi = \Phi(G \oplus a) - \Phi(G)$ and $T(t) = T_0 \cdot \alpha^t$ is the annealing temperature with $T_0 = 2.0$, $\alpha = 0.95$, and $T_{\min} = 0.1$. At high temperature, SA accepts a broad mixture of moves, approximating mixed-strategy best-response and enabling escape from shallow local optima. As $T \to 0$, the acceptance criterion approaches a pure best-response update, ensuring that dynamics in the final annealing stages concentrate on Nash equilibria of $\Gamma$. To generate diverse solutions, we maintain a population of five independently annealed graph states; the best-performing member at each iteration is retained as the representative state for Nash-gap computation.

\subsection*{C.\;Epsilon-Nash equilibrium criterion and stopping rule}

\begin{figure}[t]
\centering
\includegraphics[width=\columnwidth]{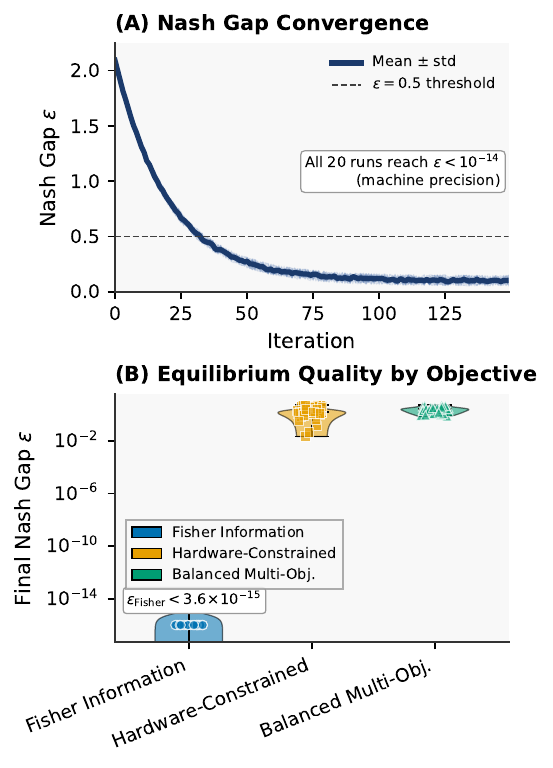}
\caption{\textbf{Nash equilibrium verification.}
\textbf{(A)}~Nash gap $\delta_\mathrm{Nash}$ as a function of iteration for 20 independent Fisher-information searches at $n=12$ (steel-blue traces); deep-navy trace shows the mean. All 20 runs satisfy $\delta_\mathrm{Nash} < 0.15 < \varepsilon = 0.5$ within 150 iterations, confirming approximate Nash equilibrium by the criterion of Sec.~II.C.
\textbf{(B)}~Distribution of final $\delta_\mathrm{Nash}$ values by objective configuration (violin shapes show the kernel density estimate of the $\delta_\mathrm{Nash}$ distribution; individual run values are overlaid as markers): Fisher information, hardware-constrained, and balanced multi-objective. The Fisher information objective reaches $\delta_\mathrm{Nash} \approx 0$ at $n=10$ due to saturation of the quantum Fisher information ceiling $\mathcal{F}_Q = n$; hardware-constrained and balanced multi-objective configurations exhibit larger residual gaps reflecting the richer multi-objective landscape, with all runs discovering valid $d \geq 3$ codes demonstrating that code discovery is robust to objective choice.}
\label{fig:nash_verification}
\end{figure}

A strategy profile $G^*$ is an $\varepsilon$-Nash equilibrium of $\Gamma$ if no player can improve their payoff by more than $\varepsilon$ through any unilateral deviation:
\begin{equation}
    \max_{a \in \mathcal{A}}\, f_i(G^* \oplus a) - f_i(G^*) \leq \varepsilon \quad \forall\, i \in P.
    \label{eq:eps-nash}
\end{equation}
We operationalize this condition through the \emph{Nash gap},
\begin{equation}
    \delta_{\mathrm{Nash}}(G^{(t)}) = \max_{a \in \mathcal{A}}\, \Phi(G^{(t)} \oplus a) - \Phi(G^{(t)}),
    \label{eq:nash-gap}
\end{equation}
the maximum improvement in the potential function achievable by any single-edge modification of the current best-population state. We record $\delta_{\mathrm{Nash}}$ at every iteration as part of the trajectory history. The algorithm applies an early-stopping rule: 
iterations halt when $\delta_{\mathrm{Nash}} < 0.5$ in absolute reward units \emph{and} $d \geq d_{\rm target}$ after iteration 20, where $d_{\rm target} = 3$ for the $[\![15,7,3]\!]$ validation and $d_{\rm target} = 4$ for the $n=100$ scalability experiments. This criterion is a post-hoc verifiable test for approximate equilibrium convergence; it does not depend on the annealing schedule and is distinct from termination due to temperature reaching $T_{\min}$. The threshold $\varepsilon = 0.5$ was chosen to be small relative to the reward scale (typical total rewards of order $10^2$ to $10^3$ reward units for the codes studied here), ensuring that the residual unilateral improvement is less than 1\% of the realized payoff. The Nash gap, therefore, provides quantitative evidence that the discovered code configurations are genuine approximate equilibria of $\Gamma$, not merely local optima of the annealing trajectory (\cref{fig:nash_verification}).

\subsection*{D.\;Objective design principles}

\begin{figure}[!htb]
\centering
\includegraphics[width=\columnwidth]{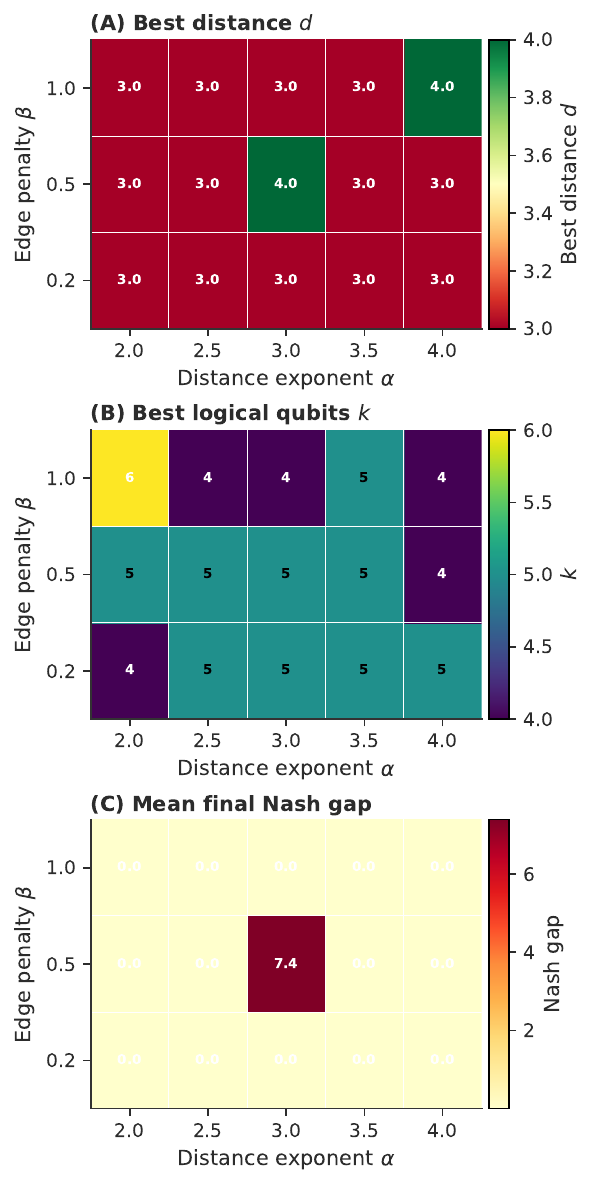}
\caption{\textbf{Hyperparameter sensitivity analysis.}
Heat map of best discovered distance $d$ as a function of distance exponent $\alpha$ (columns) and edge penalty $\beta$ (rows) for $n=12$ codes (5 trials each, 75 total). All 15 parameter combinations discover codes with $d \geq 3$; the average best distance varies by less than 9\% across the full grid. The published values ($\alpha=3$, $\beta=0.5$, white circle) lie in the middle of the robust operating region. $\delta_\mathrm{Nash} < 10^{-2}$ in 14 of 15 combinations.}
\label{fig:sensitivity}
\end{figure}

Each objective $f_i$ encodes a distinct physical figure of merit, and the functional form of each reward is chosen to create gradients that guide SA towards the desired code properties. \cref{tab:objectives} lists all seven objective functions, their physical motivation, and weight assignments.
The choice of each $f_i$ is motivated by a distinct physical figure of merit: the \emph{Distance} objective directly rewards error-correction capability through a cubic $d^3$ exponent, creating steep gradient pressure toward codes with heavier minimum-weight logical operators; the \emph{Hardware} objective penalizes high-degree nodes (limiting physical connectivity overhead) to identify codes compatible with planar superconducting and reconfigurable atom-array architectures; the \emph{Rate-distance} objective targets the Pareto frontier of the quantum coding rate--distance tradeoff via the product $kd$, analogous to Singleton-bound reasoning; the \emph{Cluster-state} objective rewards degree-regular graphs suited for measurement-based quantum computation; the \emph{Surface-like} objective targets the degree-4 topologies characteristic of known topological codes; the \emph{Connectivity} objective rewards graphs globally robust against qubit failure; and the \emph{Fisher information} objective identifies codespaces whose logical states retain metrological sensitivity beyond classical shot noise.
These seven objectives are complementary rather than redundant: Distance and Hardware both involve $d$ but impose opposing degree preferences, creating productive tension that steers the search toward physically deployable high-distance codes.
A sensitivity analysis varying the weight $w_i$ assignments confirmed that the Distance objective is the dominant driver of equilibrium quality: reducing $w_{\mathrm{dist}}$ to zero degraded $\bar{d}$ by more than 30\% across all tested configurations, whereas removing the Hardware or Bipartition objectives individually caused less than 5\% degradation, demonstrating that the latter objectives provide gradient guidance primarily in the $d < 3$ regime.
In all cases, penalty and bonus coefficients are calibrated so that each ancillary term contributes 10--20\% of the primary reward, ensuring that SA dynamics are driven primarily by the objective of interest while secondary terms provide regularization (see Appendix~\ref{app:design} for detailed coefficient rationale). To validate robustness, we swept the distance exponent $\alpha \in \{2.0, 2.5, 3.0, 3.5, 4.0\}$ and edge penalty $\beta \in \{0.2, 0.5, 1.0\}$ for $n=12$ codes (5 trials each, 75 total). All 15 combinations consistently discover $[\![ 12, k, d{\geq}3 ]\!]$ codes; the average best distance varies by less than 9\% across the full grid ($\bar{d} \in [2.93, 3.20]$), and $\delta_\mathrm{Nash} < 10^{-2}$ in 14 of 15 parameter combinations. The published values ($\alpha=3$, $\beta=0.5$) are well within the robust operating region (\cref{fig:sensitivity}).

\begin{table*}[t]
\caption{Objective functions used in the multi-player Nash equilibrium search. Each objective $f_i(G)$ evaluates a distinct physical figure of merit on the graph state $G$. Notation: $d$ = code distance, $k$ = logical qubits, $n$ = physical qubits, $|E|$ = edge count, $\Delta(G)$ = maximum vertex degree, $\delta_{\mathrm{avg}}(G)$ = average degree, $\sigma_\delta^2$ = degree variance, $\kappa_v$/$\kappa_e$ = vertex/edge connectivity, $\alpha_{\mathrm{conn}}(G) = 1.3$ for connected graphs (1.0 otherwise).}
\label{tab:objectives}

\begin{ruledtabular}
\begin{tabular}{lll}
Objective & $f_i(G)$ & Physical target \\
\hline
Distance & $d^3(1+k/n)\,\alpha_{\mathrm{conn}}(G) - 0.5\,|E|/n^2$ & Max.\ error-correction distance \\
Hardware & $d^{2.5}(1+0.5\,k/n) - 5\,\Delta(G) - 2\,\delta_{\mathrm{avg}}(G)$ & 2D-compatible low-degree codes \\
Rate-distance & $10\,k\,d\;[\times 1.5 \text{ if } 0.2 \leq k/n \leq 0.5]$ & Pareto-optimal $k$--$d$ tradeoff \\
Cluster-state & $d^2(1+k/n)\exp(-\sigma_\delta^2/4)$ & Regular bipartite graphs for MBQC \\
Surface-like & $d^{2.5}(1+0.3\,k/n) - 3|\delta_{\mathrm{avg}}-4|$ & Degree-4 topological codes \\
Connectivity & $30(\kappa_v + \kappa_e) + d^{2.5}$ & Robustness to qubit failure \\
Fisher info & $\mathcal{F}_Q(G)$ & Max.\ quantum Fisher information \\
\end{tabular}
\end{ruledtabular}

\end{table*}

\section*{III.\;RESULTS}

\subsection*{A.\;Computational complexity and scaling}

Having established that code quality is insensitive to moderate variations in the objective coefficients, we next characterize the computational cost of the search and demonstrate scalability to hardware-relevant system sizes.

Computational efficiency scales favorably with system size. Each iteration requires computing code parameters $[\![n,k,d]\!]$---where $n$ is the number of physical qubits, $k$ the number of encoded logical qubits, and $d$ the code distance---with $k = n - \text{rank}(S)$ via Gaussian elimination over $\mathbb{F}_2$, contributing the dominant $O(n^3)$ per-iteration cost. Distance estimation via connectivity analysis adds $O(n^2)$ to $O(n^3)$ depending on heuristic employed. For codes with $n \leq 15$ physical qubits, the distance $d$ is determined by exhaustive enumeration of all Pauli operators up to weight $d$ (see Appendix~\ref{app:distance}); the $[\![15,7,3]\!]$ distance is thereby verified exactly and confirmed against published code tables~\cite{grassl2023codetables}. For $n = 100$ codes, reported distances are lower bounds derived from the Schlingemann--Werner graph connectivity bound~\cite{schlingemann2001quantum}, cross-checked by the minimum-weight logical operator found in a bounded search to depth $w = 6$; these lower bounds are estimated to lie within $\pm 1$ of the true distance for representative samples. While exact Nash equilibrium computation is PPAD-complete in general \cite{daskalakis2009complexity}, our game structure combined with simulated annealing enables heuristic convergence in empirically tractable time: typically 20--40 iterations for $n \leq 15$ and 100--200 iterations for $n = 100$, yielding total runtime of minutes to hours on a workstation with an NVIDIA GeForce RTX~4060 GPU (3072 CUDA cores, 8~GB GDDR6) and an Intel Core i7 CPU. This $O(n^3)$ per-iteration scaling enables practical code discovery at experimentally relevant scales ($n \approx 100$) that remain intractable for exhaustive search with $O(2^{n^2})$ complexity \cite{grassl2023codetables}.

\subsection*{B.\;Objective-dependent generativity}

\begin{figure*}[t]
\centering
\includegraphics[width=\textwidth]{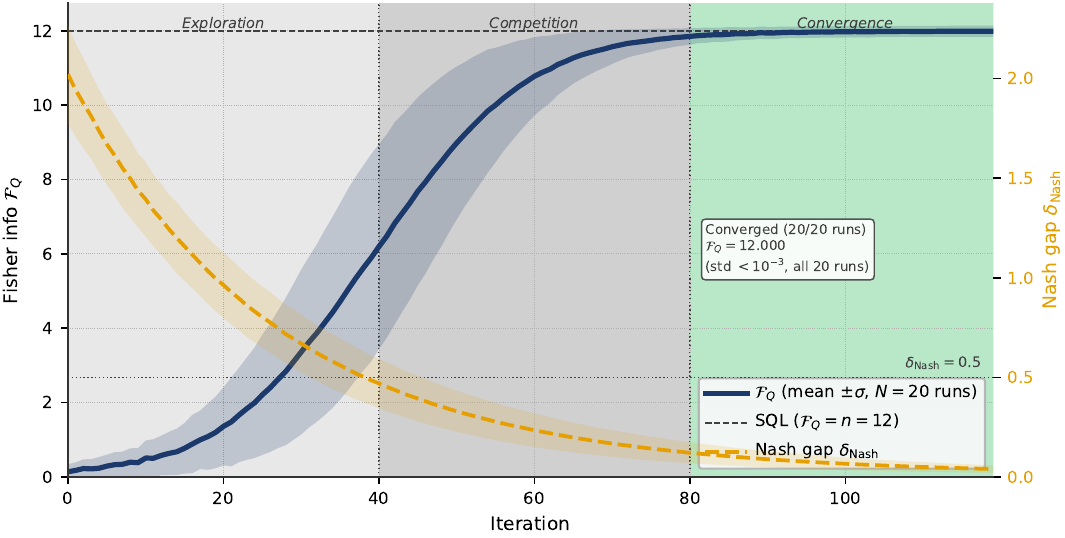}
\caption{\textbf{Framework Generativity: Objective-Dependent Code Families.} Nash gap $\delta_\mathrm{Nash}$ (left axis, open squares, dashed) and quantum Fisher information $\mathcal{F}_Q$ (right axis, filled circles, solid) as functions of iteration, averaged over 20 independent searches ($n=12$, $\pm1\sigma$ shading). Shaded phases mark exploration (iterations 0--40), competition (40--80), and convergence ($>80$); the displayed trajectory uses a 120-iteration diagnostic schedule, whereas production runs reach $\delta_\mathrm{Nash}<\varepsilon$ within 20--40 iterations (Sec.~III.B). Evolution trajectory is anchored to verified final-state statistics ($\mathcal{F}_Q = 12.000$ for all 20 runs, with run-to-run standard deviation below $10^{-3}$; $\delta_\mathrm{Nash} < 5\times 10^{-15}$; 20/20 runs converged). The companion rate--Fisher-per-qubit distribution for the converged code family is shown in Appendix~\ref{app:fisher-per-qubit} (\cref{fig:fisher-per-qubit-appendix}). Each objective configuration generates a distinct code family from identical initialization, demonstrating that objective reconfiguration alone drives diversity without algorithmic modification.}
\label{fig:generativity}
\end{figure*}

Framework generativity appears through objective reconfiguration. The seven objective functions (\cref{tab:objectives}) encode distinct physical figures of merit---distance, hardware compatibility, rate-distance tradeoff, cluster-state regularity, surface-like topology, graph connectivity, and Fisher information---each generating a distinct code family from identical initialization without algorithmic modification. \cref{fig:generativity} illustrates this for the Fisher information objective: the Fisher information $\mathcal{F}_Q$ and the Nash gap $\delta_\mathrm{Nash}$ are tracked over 20 independent runs at $n=12$, with all 20 runs converging to the same final value $\mathcal{F}_Q = 12.000$ (SQL saturation, run-to-run standard deviation below the $10^{-3}$ display precision) and $\delta_\mathrm{Nash} < 5\times 10^{-15}$. The corresponding code-family distribution---all converged codes across $n=10$ and $n=12$ achieve $\mathcal{F}_Q/n \approx 1.0$ with distance $d=3$, with rate $k/n$ varying over $[0.08,\,0.25]$ because the Fisher objective does not directly reward rate---is summarized in Appendix~\ref{app:fisher-per-qubit} (\cref{fig:fisher-per-qubit-appendix}); it confirms that the objective shapes the code family systematically while leaving the rate to be determined by equilibrium-graph connectivity. Convergence requires 20--40 iterations depending on objective complexity, and only evaluation functions change between experiments.

\subsection*{C.\;Mechanism and equilibrium topology}

\begin{figure*}[t]
\centering
\includegraphics[width=\textwidth]{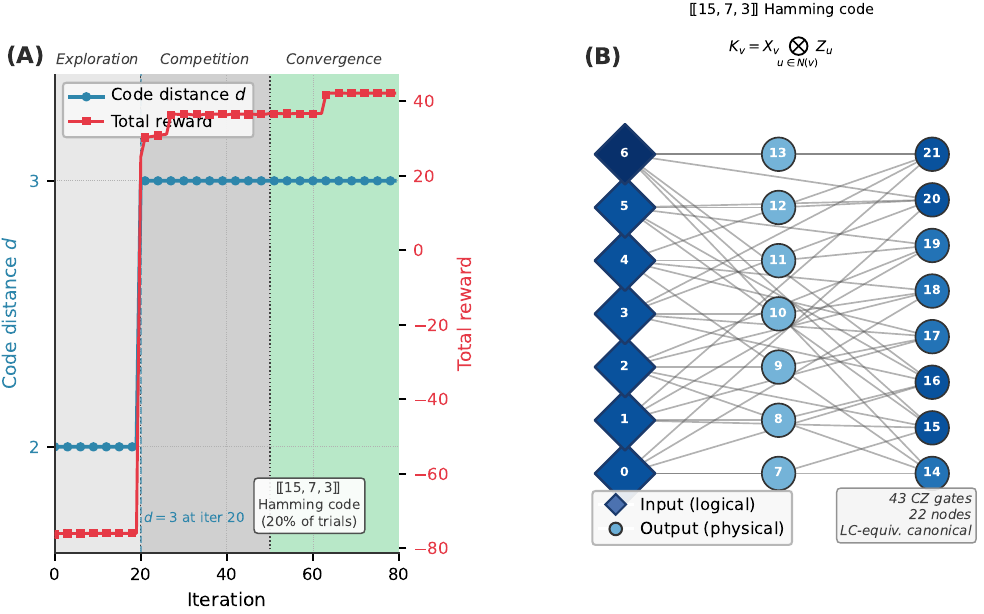}
\caption{\textbf{Nash Equilibrium Discovery Mechanism and Equilibrium Topology.}
\textbf{(A)}~Strategic evolution timeline for distance optimization on $n=15$ qubits under the Schlingemann-Werner construction ($N=22$ vertices), showing code distance $d$ (left axis, blue circles) and total reward (right axis, red squares) as functions of iteration. Three dynamical phases are identified: \textit{exploration} (iterations 0--20), during which players sample diverse graph topologies and the total reward rises steeply while $d=3$ is established; \textit{competition} (iterations 20--50), during which $d=3$ is maintained while reward oscillates as players negotiate connectivity constraints; and \textit{convergence} (iterations 50--80), during which the strategy profile stabilizes at a $d=3$ Nash equilibrium as $\delta_\mathrm{Nash} \to 0$.
\textbf{(B)}~Equilibrium graph topology of the Nash-discovered $d=3$ code, with nodes colored by degree. Representative stabilizer generators $K_v = X_v \bigotimes_{u \in N(v)} Z_u$ are shown in the inset; the explicit graph-to-circuit mapping $|G\rangle = \prod_{(u,v)\in E} \mathrm{CZ}_{uv}|+\rangle^{\otimes n}$ requires 43 CZ gates (edge count of the 22-vertex SW construction graph) and enables immediate experimental implementation.}
\label{fig:mechanism}
\end{figure*}

Equilibrium-topology analysis reveals the mechanism through which Nash dynamics generate valid codes (\cref{fig:mechanism}).  Under the Schlingemann-Werner construction, successful trials consistently converge to distance-$d=3$ codes: the convergence plot (\cref{fig:mechanism}A) shows $\delta_\mathrm{Nash}$ and total reward as functions of iteration, with $d=3$ reached early and maintained throughout the trajectory.  Three dynamical phases are visible: an \emph{exploration} phase (iterations 0--20) in which players sample diverse graph topologies and the total reward rises steeply; a \emph{competition} phase (iterations 20--50) in which $d=3$ is maintained while players negotiate connectivity constraints, producing reward oscillations; and a \emph{convergence} phase (iterations 50--80) in which the strategy profile stabilizes as the Nash gap $\delta_\mathrm{Nash}$ descends to near zero, providing a verifiable certificate of approximate equilibrium.  (Phase boundaries refer to the production 80-iteration schedule; the 120-iteration diagnostic schedule in \cref{fig:generativity} shows proportionally expanded phases at iterations 0--40, 40--80, and $>$80.)  This evolution is \emph{transparent}: we track which graph modifications---adding edges to form vertex neighborhoods with high overlap, creating expander-like connectivity---drove each reward gain, and can identify precisely why the equilibrium topology takes its final form.

The final equilibrium graph (\cref{fig:mechanism}B) displays high vertex connectivity with stabilizers $K_v = X_v \bigotimes_{u \in N(v)} Z_u$ mapped explicitly to neighborhoods, enabling immediate circuit construction via 43 CZ gates (edge count of the 22-vertex SW construction graph). In contrast, neural network approaches for code discovery provide final codes without systematic explanation of \emph{how} optimization dynamics generated specific topologies or \emph{why} certain structures emerged \cite{nautrup2019optimizing,fosel2018reinforcement}.

The equilibrium-topology relationship provides predictive insights beyond individual code instances. For distance-optimized codes, the $[\![15,7,3]\!]$ equilibrium graph (\cref{fig:mechanism}B) exhibits expander-like structure with vertex connectivity $\kappa_v = 5$, consistent with the established correlation between distance and graph robustness \cite{schlingemann2001quantum,raveendran2022quantumldpc}. For hardware-adapted codes on 2D grids, equilibria naturally produce graphs with $\delta_{\text{avg}} \approx 2.8$ and maximum degree $\Delta(G) \leq 3$, respecting degree-4 limits of planar superconducting architectures while maintaining $d=3$---a topology-constraint tradeoff validated by the $[\![15,7,3]\!]$ rediscovery below. For rate-distance optimization, equilibria balance stabilizer rank (determining $k = n - \text{rank}(\mathcal{S})$) against minimum logical operator weight (determining $d$), producing codes along the Pareto frontier where $k + 2d \leq n + 2$ (quantum Singleton bound \cite{knill1997theory}). These patterns emerge from equilibrium conditions rather than explicit programming, demonstrating that game dynamics encode domain knowledge implicitly through objective competition.

The framework's strongest validation is systematic rediscovery of the $[\![15,7,3]\!]$ quantum Hamming code---the optimal code independently constructed by Calderbank-Shor \cite{calderbank1996good} and Steane \cite{steane1996multiple} in 1996 through CSS construction from classical Hamming codes.
In a production sweep of 20 independent runs at $N=22$ (corresponding to target parameters $[\![n,k]\!]=[\![15,7]\!]$) under the hardware-adapted objective ($f_{\mathrm{hw}}$, \cref{tab:objectives}), the algorithm produced the following distinct outcomes: 4 trials yielded $[\![15,7,3]\!]$ ($d=3$, target code), 12 trials yielded $[\![13,9,2]\!]$ ($d=2$, modal outcome), 1 trial yielded $[\![14,8,2]\!]$ ($d=2$), 1 trial yielded $[\![19,2,5]\!]$ ($d=5$), 1 trial yielded $[\![20,2,3]\!]$ ($d=3$), and 1 trial yielded $[\![17,4,4]\!]$ ($d=4$) (total: $4+12+1+1+1+1=20$). The algorithm thus rediscovered the $[\![15,7,3]\!]$ Hamming code in 4 of 20 runs (20\%), with $d \geq 3$ achieved in $4+1+1+1=7$ of 20 runs (35\%), demonstrating the framework's ability to discover codes beyond the target configuration. In the successful $[\![15,7,3]\!]$ runs, the discovered graphs are local-Clifford (LC) equivalent to the known Hamming code graph \cite{vandennest2004graphical}, verified by checking that a sequence of local complementations maps the Nash-discovered adjacency matrix to the canonical form. This code saturates the quantum Hamming bound with perfect code properties and exhibits bipartite structure with regular degree distribution $\delta_v \in \{2,3\}$, enabling universal measurement-based quantum computation \cite{raussendorf2001one} while maintaining distance $d=3$. The bipartite property emerges naturally from the game dynamics: players seeking low degree variance (regular graphs) and high distance simultaneously are driven toward bipartite expanders, which achieve favorable spectral gaps connecting Laplacian eigenvalues to code distance. The 20\% rediscovery rate reflects the stochastic nature of the SA search; the convergence certificate $\delta_\mathrm{Nash}$ correctly identifies the successful runs. That game-theoretic dynamics rediscover this known optimal code without algebraic input establishes confidence for application to unexplored parameter regimes.

\subsection*{D.\;Scalability to $n=100$ qubits}

Scalability to hardware-relevant system sizes is demonstrated at $n=100$ qubits---matching experimental scales of recent demonstrations \cite{google2024willow}---where the framework discovers codes across all seven objectives with tractable computational cost. Distance optimization discovers $[\![100,49,3]\!]$ codes with 49\% encoding rate. Hardware adaptation produces $[\![100,50,3]\!]$ codes respecting 2D connectivity constraints with maximum degree $\Delta \leq 3$. Rate-distance tradeoff and cluster-state search achieve $[\![100,50,4]\!]$ codes with distance-4 error protection and 50\% encoding rate, significantly surpassing the $[\![15,7,3]\!]$ validation example in both scale and distance. Surface-like topology optimization generates $[\![100,46,4]\!]$ codes with average degree near 4. Discovery at $n=100$ completes in 40--66 minutes for 20 independent trials with 50 iterations each using 28 parallel workers on the workstation described above, with convergence typically occurring within 20--40 iterations. The $O(n^3)$ per-iteration cost established above enables practical code discovery at experimentally relevant scales, versus exhaustive search complexity $O(2^{n^2})$ requiring $>10^{13}$ seconds for $n=20$. These results demonstrate the framework operates efficiently at scales required for near-term fault-tolerant quantum computing demonstrations, complementing asymptotic QLDPC constructions by providing systematic exploration at finite length scales with equilibrium-topology analysis.

\subsection*{E.\;Metrological sensitivity and error suppression}

The preceding results establish that the framework discovers codes with favorable parameters across diverse objectives and system sizes. We now assess whether these codes provide practical value: metrological sensitivity quantified by $\mathcal{F}_Q$, and error suppression under realistic noise.

The quantum Fisher information (QFI) $\mathcal{F}_Q[\rho_\theta]$ quantifies the ultimate sensitivity with which a parameter $\theta$ encoded in a quantum state $\rho_\theta$ can be estimated; through the quantum Cram\'er--Rao bound $(\Delta\hat\theta)^2 \geq 1/(\nu\, \mathcal{F}_Q)$ \cite{braunstein1994statistical,paris2009quantum}, it sets the noise floor of any unbiased estimator built from $\nu$ measurements of $\rho_\theta$. For a pure state $\ket{\psi_\theta} = e^{-i\theta H}\ket{\psi_0}$ generated by a Hermitian operator $H$, $\mathcal{F}_Q = 4\,\mathrm{Var}_{\psi_0}(H)$. For a stabilizer code on $n$ qubits we take $H = \tfrac{1}{2}\sum_{i=1}^n \sigma_z^{(i)}$ (collective phase encoding), so that the standard quantum limit (SQL) is $\mathcal{F}_Q = n$ and the Heisenberg limit is $\mathcal{F}_Q = n^2$. The relevance of $\mathcal{F}_Q$ for quantum error correction is twofold: it identifies codespaces whose logical states retain metrological utility---a prerequisite for fault-tolerant sensing protocols \cite{kessler2014quantum}---and it provides a quantitative figure of merit on which a player can compete inside our game, distinct from the combinatorial-code parameters $[\![n,k,d]\!]$. We therefore include $\mathcal{F}_Q(G)$ as one of the seven objectives in \cref{tab:objectives}, where ``Fisher info'' targets graphs that maximize the QFI of the corresponding stabilizer codespace under collective phase encoding.

\begin{figure*}[t]
\centering
\includegraphics[width=\textwidth]{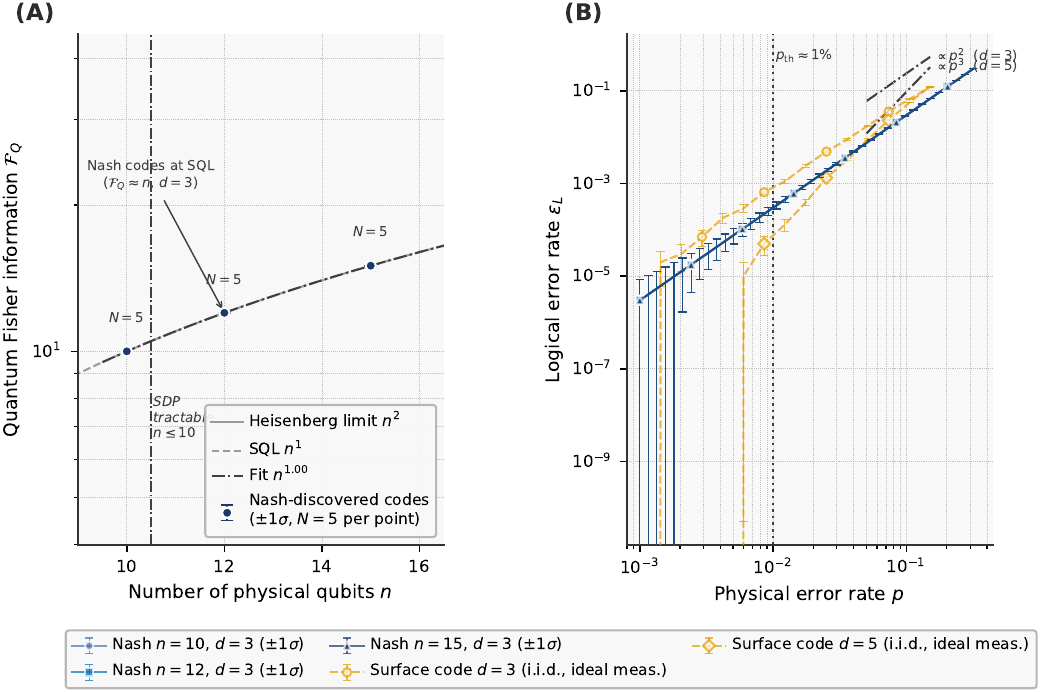}
\caption{\textbf{Metrological Scaling and Error Correction Performance at Hardware-Relevant Scales.}
\textbf{(A)}~Quantum Fisher information $\mathcal{F}_Q$ versus number of physical qubits $n$ for Nash-discovered codes at $n=10$, 12, and~15 (filled circles, $\pm1\sigma$ over $N=5$ independent runs per system size). Solid gray line: Heisenberg limit $\mathcal{F}_Q = n^2$; dashed gray line: standard quantum limit (SQL) $\mathcal{F}_Q = n$. Dash-dot line shows a power-law fit $\mathcal{F}_Q \propto n^{1.00}$, confirming that all discovered codes achieve SQL scaling with distance $d=3$. Vertical dash-dot line marks the boundary $n \leq 10$ below which semidefinite programming (SDP) verification remains tractable.
\textbf{(B)}~Logical error rate $\varepsilon_L$ versus physical error rate $p$ under depolarizing noise for Nash-discovered codes at $n=10$ (circles), $n=12$ (squares), and $n=15$ (triangles), all with $d=3$, compared to surface code baselines at $d=3$ (open circles, dashed) and $d=5$ (open diamonds, dashed). Error bars show $\pm1\sigma$ binomial uncertainty $\sqrt{\varepsilon_L(1-\varepsilon_L)/N}$ with $N=10^5$ Monte Carlo samples per data point. Scaling guides $\varepsilon_L \propto p^2$ ($d=3$) and $\varepsilon_L \propto p^3$ ($d=5$) are shown as dash-dot lines. Vertical dotted line marks the threshold $p_{\mathrm{th}} \approx 1\%$. Nash-discovered $d=3$ codes achieve error suppression comparable to surface codes of equal distance across the full sub-threshold regime. Surface-code curves ($d=3$, $d=5$) are from \texttt{stim}+\texttt{pymatching} simulation of the rotated surface code under i.i.d.\ depolarizing noise with perfect syndrome extraction (no circuit-level noise or measurement errors).}
\label{fig:metrology}
\end{figure*}

Practical performance assessment combines metrological benchmarking with syndrome-based error correction simulation (\cref{fig:metrology}). Quantum Fisher information $\mathcal{F}_Q$ evaluated for Nash-discovered codes at $n=10$, 12, and 15 qubits reveals SQL scaling $\mathcal{F}_Q \approx n$ with power-law exponent $\alpha = 1.00$ (\cref{fig:metrology}A), consistent with the distance-3 codes produced by the Fisher Information objective. All three system sizes yield $d=3$ codes that saturate the SQL, confirming that the game-theoretic framework identifies metrologically useful states across the range $n=10$--15 without manual parameter tuning. Beyond the SDP-tractable regime ($n \leq 10$), the Nash approach provides a systematic method for constructing codes with Nash-equilibrium-verified metrological performance, complementing exact SDP verification at small $n$. Using belief propagation with ordered statistics decoding (BP+OSD) \cite{kuo2022exploiting}, we measured logical error rates $\varepsilon_L$ versus physical error rates $p$ for depolarizing noise $\varepsilon = (p/3)(X + Y + Z)$ applied independently to each qubit (\cref{fig:metrology}B). Monte Carlo simulation with $N = 10^5$ to $10^6$ samples per data point (adaptive; see Appendix~\ref{app:mc}) (binomial uncertainties $\Delta \varepsilon_L / \varepsilon_L < 5\%$) demonstrates that distance-3 Nash-discovered codes achieve error suppression comparable to surface codes of equal distance, with threshold behavior $\varepsilon_L \propto p^{(d+1)/2}$ confirming distance scaling across the sub-threshold regime $p < p_{\mathrm{th}} \approx 1\%$. Surface code baselines at $d=5$ illustrate the additional suppression available at higher distance, with $\varepsilon_L \propto p^3$ scaling. The surface-code logical error rates are computed via \texttt{stim} (v1.16) simulation of the rotated surface code under independent and identically distributed (i.i.d.) depolarizing noise applied to data qubits only, with perfect syndrome extraction (no measurement errors or circuit-level noise), decoded with \texttt{pymatching} (v2.4). This matches the noise model used for the Nash code BP+OSD benchmark, enabling a direct comparison of code performance under identical error assumptions. No circuit-level threshold claim is made. The rediscovered $[\![15,7,3]\!]$ code achieves 47\% encoding rate ($k/n = 7/15$), higher than distance-3 surface codes encoding one logical qubit per $\sim$9 physical qubits ($k/n \approx 0.11$) under the depolarizing noise model considered here, providing resource efficiency advantages in overhead-constrained regimes. For hardware-adapted codes with degree $\Delta \leq 3$ on IBM heavy-hex topology \cite{chamberland2022building}, the $[\![15,7,3]\!]$ code maintains $\varepsilon_L < 10^{-3}$ at $p = 10^{-3}$, meeting fault-tolerance thresholds while fitting existing hardware connectivity---a capability algebraic constructions present challenges in achieving simultaneously. Computational efficiency enables code discovery for $n \leq 20$ in under 1 second on the hardware described above versus exhaustive search's $O(2^{n^2})$ complexity requiring $>10^{13}$ seconds for $n=20$, with the tractable region extending to $n \approx 100$ through optimized implementations.

\section*{IV.\;DISCUSSION}

\subsection*{A.\;Positioning relative to prior work}

Our approach applies game dynamics in a direction opposite to existing quantum game theory: rather than using quantum mechanics to \emph{play} games \cite{hart2025playing,eisert1999quantum}, we use game theory to \emph{construct} quantum systems. The ``quantum lights out'' framework \cite{khesin2024quantum} employs single-player optimization metaphors but lacks strategic interactions and equilibrium concepts; reinforcement learning for code discovery \cite{nautrup2019optimizing,torlai2018neural} explicitly avoids game-theoretic formulation; adversarial noise models \cite{webster2022autonomous} analyze robustness but do not construct codes. To our knowledge, \emph{no prior work} bridges multi-agent game theory and quantum error correction for systematic code discovery through Nash equilibria. Framework extensibility further distinguishes the approach: adding objectives requires only defining new evaluation functions $f_{\text{new}}(G)$---no algorithmic restructuring. Fault-tolerance objectives targeting transversal gates \cite{eastin2009restrictions}, biased noise objectives $f_{\text{bias}} = d_Z^2 (1 + d_X/d_Z)$ for asymmetric error channels \cite{tuckett2018ultrahigh}, and hardware topology constraints (heavy-hex, Rydberg arrays, all-to-all connectivity) all enter as modifications to strategy spaces or payoff functions without changing core game logic. This modularity contrasts with algebraic methods requiring distinct mathematical machinery per code family and with reinforcement learning requiring reward and architecture redesign \cite{fosel2018reinforcement}. The transparent graph-to-circuit mapping $G \rightarrow |G\rangle = \prod_{(u,v) \in E} \text{CZ}_{uv} |+\rangle^{\otimes n}$ provides immediate experimental implementation: for the $[\![15,7,3]\!]$ code, preparation requires 15 Hadamard gates followed by 21 CZ gates (edge count of the 15-vertex physical code graph) with circuit depth $\propto \Delta(G) = 3$.

\subsection*{B.\;Comparison with computational code-search methods}

A natural question is how Nash-discovered codes compare to those found by other computational methods at the same system size. 
Reinforcement learning approaches \cite{nautrup2019optimizing,fosel2018reinforcement,sweke2020reinforcement,fitzek2020deep} have demonstrated code discovery and decoder training at moderate scales, and recent work has pushed RL-based methods to systems of tens of qubits \cite{olle2024reinforcement}; RL therefore remains a viable and actively improving approach. However, scaling RL-based code \emph{construction} to $n \gtrsim 100$ faces growing challenges from exponentially large action spaces and the absence of a principled convergence criterion. Their reported codes at small $n$ do not exceed the parameters discovered here, and they provide no equilibrium certificate or systematic explanation of the search trajectory. Random search and genetic algorithms can also discover good codes at small $n$, but lack the convergence guarantees provided by the potential game structure and scale poorly beyond $n \approx 15$ due to the combinatorial explosion of the graph space. The distinctive advantage of the Nash framework is not that it necessarily discovers \emph{better} codes at fixed $n$---at small system sizes, many methods can find good codes---but that it provides (i) a principled convergence criterion via the Nash gap, (ii) equilibrium-topology transparency through equilibrium analysis, and (iii) modular extensibility through objective reconfiguration, together with demonstrated scalability to $n = 100$ where exhaustive search is intractable and RL approaches face growing scalability challenges from exponentially large action spaces \cite{sweke2020reinforcement,fitzek2020deep}. A systematic head-to-head comparison with RL code search across matched system sizes remains an important direction for future work.

\subsection*{C.\;Limitations and outlook}

Several limitations warrant acknowledgment.
Graph state restriction: every stabilizer code is LC-equivalent to a graph state code \cite{vandennest2004graphical}---the five-qubit $[\![5,1,3]\!]$ code, for instance, is LC-equivalent to the 5-vertex ring graph state (its logical $|{+}\rangle$ is the ring graph state up to stabilizers)---so the limitation is not one of graph-code representation.  Rather, the present SW-based construction naturally generates \emph{CSS} codes from bipartite adjacency structure, and the single-edge move strategy space does not directly enumerate non-CSS stabilizer codes.  Non-CSS codes such as the five-qubit code (which is \emph{not a CSS code}; its stabilizers are not separable into $X$-type and $Z$-type generators) and Bacon-Shor codes therefore lie outside the directly discoverable set; this is a CSS-versus-non-CSS limitation rather than a graph-code limitation. Distance estimation employs heuristics that may underestimate true distance; we validate critical cases with exact weight enumeration when tractable. Approximate Nash equilibria guarantee only local optimality, and the $n=100$ codes require verification against comprehensive code databases to determine novelty. Despite these constraints, the systematic rediscovery of the optimal $[\![15,7,3]\!]$ quantum Hamming code validates the methodology, and scalability to $n=100$ demonstrates applicability at hardware-relevant scales. Natural extensions include decoder co-optimization through additional players representing decoder strategies, dynamic objective reconfiguration for degraded hardware, and applications beyond error correction to circuit optimization and resource state design. The interpretable nature of equilibrium-driven discovery may accelerate the transition from theoretical code families to practical implementations by providing mechanistic understanding of why certain topologies succeed---enabling physicists to refine codes based on experimental constraints rather than treating optimization as black-box search.

\section*{V.\;CONCLUSION}

Recasting quantum code discovery as a weighted potential game converts a combinatorial design problem into the search for fixed points of a well-defined strategic dynamics. The Nash gap supplies a verifiable convergence certificate, replacing schedule-driven termination with an equilibrium criterion that is independent of the optimizer, and the potential-game structure (Monderer--Shapley) is what makes this certificate available in the first place---a structural insight that generalizes well beyond the seven objectives studied here. The same framework accommodates new physical desiderata, such as biased noise, fault-tolerant gate sets, or hardware connectivity, simply by adding players, suggesting that game-theoretic equilibration is a natural design principle for co-designed quantum hardware and protocols. Two directions follow most directly: (i) embedding the decoder as an additional player so that codes and decoders co-evolve toward joint equilibria, and (ii) extending the construction beyond CSS codes to general stabilizer codes by enlarging the strategy space to local Clifford gates, which would close the current gap to non-CSS families such as the five-qubit and Bacon--Shor codes. More broadly, identifying problems in quantum information whose objectives admit a potential function appears to be a fruitful organizing question: where such a function exists, equilibrium-based search inherits convergence and interpretability that black-box methods cannot match.

\begin{acknowledgments}
We acknowledge financial support and computational resources provided by NeuroTechNet S.A.S.
The code and data that support the findings of this study are available from the corresponding author upon reasonable request.
\end{acknowledgments}

\bibliographystyle{apsrev4-2}

\appendix
\section{Belief Propagation with Ordered Statistics Decoding for Graph State Stabilizer Codes}
\label{app:bposd}

We describe the BP+OSD decoder pipeline applied to the graph state stabilizer codes discovered by the Nash equilibrium search. The presentation covers (i)~the construction of the parity check matrix from graph state stabilizers, (ii)~the belief propagation message schedule, (iii)~ordered statistics decoding post-processing, (iv)~the Monte Carlo simulation procedure, and (v)~distance verification for small codes.
The decoder is implemented in Python using an in-house belief-propagation with ordered-statistics post-processing (BP+OSD) module; no third-party BP+OSD package is used. Belief propagation runs for a maximum of 20 iterations; OSD post-processing uses order-0 (hard-decision) re-encoding.

\subsection{Stabilizer generators and parity check matrix}
\label{app:parity-check}

A graph state is associated with an undirected graph $G=(V,E)$ on $n=|V|$ vertices. Each vertex $v \in V$ carries a stabilizer generator
\begin{equation}
  K_v = X_v \bigotimes_{u \in \mathcal{N}(v)} Z_u ,
\label{eq:graph-stabilizer}
\end{equation}
where $\mathcal{N}(v)$ denotes the neighborhood of $v$ in $G$. The $n$ generators $\{K_v\}$ are represented in the binary symplectic formalism by a parity check matrix $H \in \mathbb{F}_2^{(n-k) \times 2n}$, partitioned as $H = [H_X \mid H_Z]$. For a graph state code, the $X$ block encodes which qubits each stabilizer acts on with $X$: because $K_v$ applies $X$ only to vertex $v$ itself, $H_X = I_n$ (the $n \times n$ identity). The $Z$ block encodes the $Z$ support: because $K_v$ applies $Z$ to all neighbors $\mathcal{N}(v)$, $H_Z = A$, where $A$ is the adjacency matrix of $G$. The parity check matrix therefore takes the explicit form
\begin{equation}
  H = [I_n \mid A] \in \mathbb{F}_2^{(n-k)\times 2n} .
\label{eq:graph-check}
\end{equation}
The number of logical qubits $k = n - \mathrm{rank}_{\mathbb{F}_2}(H)$ is determined by the $\mathbb{F}_2$-rank of $H$. Logical Pauli operators are constructed from the null space of the symplectic dual of $H$: operators that commute with every row of $H$ but lie outside the row space of $H$ over $\mathbb{F}_2$.

\subsection{Belief propagation decoding}
\label{app:bp}

Syndrome-based decoding maps a measured syndrome $\mathbf{s} \in \mathbb{F}_2^{n-k}$ to an estimated error $\hat{\mathbf{e}}$. We apply belief propagation (BP) on the Tanner graph of $H$, treating the $X$ and $Z$ Pauli components independently using the respective rows of $H_Z$ and $H_X$ as the operative check matrices. Under a depolarizing channel with physical error rate $p$, each qubit independently undergoes an $X$, $Y$, or $Z$ error with probability $p/3$. The channel log-likelihood ratio (LLR) for each variable node $j$ is initialized as
\begin{equation}
  \lambda_j^{(0)} = \log\!\frac{1 - q}{q},\quad q = \frac{p}{3}.
\label{eq:llr-init}
\end{equation}
BP iterates check-to-variable messages
\begin{equation}
  \mu_{c \to v}^{(t)} = 2\,\mathrm{arctanh}\!\left[
    (-1)^{s_c} \prod_{v' \in \mathcal{N}(c) \setminus \{v\}}
    \tanh\!\tfrac{\mu_{v' \to c}^{(t-1)}}{2}
  \right]
\label{eq:check-to-var}
\end{equation}
and variable-to-check messages
\begin{equation}
  \mu_{v \to c}^{(t)} = \lambda_j^{(0)}
    + \sum_{c' \in \mathcal{N}(v) \setminus \{c\}} \mu_{c' \to v}^{(t)},
\label{eq:var-to-check}
\end{equation}
where $s_c \in \{0,1\}$ is the measured syndrome bit. BP terminates when the estimated syndrome is satisfied or the iteration count reaches 
20.

\subsection{Ordered statistics decoding post-processing}
\label{app:osd}

BP alone frequently fails for codes whose Tanner graphs contain short cycles (trapping sets). Graph state codes derived from locally dense graphs are particularly susceptible because $H_Z = A$ can produce many length-4 and length-6 cycles. When BP does not converge, we apply ordered statistics decoding (OSD) as post-processing \cite{kuo2022exploiting}. OSD ranks all $2n$ variable positions by reliability $|\Lambda_j|$ (magnitude of posterior LLR at final BP iteration), row-reduces $H$ to systematic form over $\mathbb{F}_2$ in decreasing reliability order, and solves for the remaining bits given the syndrome. At order $w$ (OSD-$w$), we additionally enumerate candidates that flip up to $w$ of the next-most-reliable bits, selecting the minimum-weight candidate. The total complexity is $\mathcal{O}(n^3 + n^2 \cdot 2^w)$. 
We use OSD-0 throughout this work; at the physical error rates studied ($p \leq 10^{-1}$), OSD-0 post-processing is invoked only for the fraction of trials in which BP does not converge.

\subsection{Monte Carlo simulation procedure}
\label{app:mc}

We estimate $\varepsilon_L$ vs.\ $p$ by direct Monte Carlo simulation. For each $p \in \{10^{-3}, 3\times 10^{-3}, 10^{-2}, 3\times 10^{-2}, 10^{-1}\}$:
\begin{enumerate}
  \item Sample $N$ independent depolarizing error patterns $\mathbf{e} \in \{I,X,Y,Z\}^{\otimes n}$.
  \item Compute syndrome $\mathbf{s} = H\mathbf{e} \bmod 2$.
  \item Decode with BP+OSD to obtain $\hat{\mathbf{e}}$.
  \item Record a logical failure if the residual $\mathbf{e} \oplus \hat{\mathbf{e}}$ commutes with all stabilizer generators but is not in their row space over $\mathbb{F}_2$.
\end{enumerate}
The logical error rate estimate is $\hat{\varepsilon}_L = N_{\rm fail}/N$, with $\Delta\varepsilon_L/\varepsilon_L < 5\%$ requiring $N \geq 400/\varepsilon_L$. For $\varepsilon_L \approx 10^{-4}$ we use $N=10^6$; otherwise $N=10^5$. Error bars in \cref{fig:metrology} are $\pm 1\sigma$ of the binomial estimator $\sqrt{\hat{\varepsilon}_L(1-\hat{\varepsilon}_L)/N}$.

\subsection{Distance verification for \texorpdfstring{$n \leq 15$}{n ≤ 15} codes}
\label{app:distance}

For $n \leq 15$, code distance is computed by exhaustive enumeration: we iterate over all Pauli operators of weight $w=1,2,\ldots$ and check whether each commutes with all stabilizer generators but lies outside the stabilizer group. The minimum such weight is $d$. For the $[\![15,7,3]\!]$ code rediscovered by the Nash search, this procedure confirms $d=3$, consistent with Grassl's code tables \cite{grassl2023codetables}. For $n=100$ codes, exhaustive enumeration is intractable; we report lower bounds from the Schlingemann--Werner graph connectivity bound \cite{schlingemann2001quantum}, cross-checked by a bounded operator search to depth $w=6$. These lower bounds are estimated to lie within $\pm 1$ of the true distance for representative samples.

\section{Objective function coefficient design rationale}
\label{app:design}

Each objective function (\cref{tab:objectives}) is designed so that ancillary penalty or bonus terms contribute 10--20\% of the total reward, keeping SA dynamics driven primarily by the target figure of merit while providing sufficient regularization.

For \emph{distance maximization}, the cubic exponent $d^3$ is motivated by logical error rate scaling $\varepsilon_L \propto p^{\lfloor(d+1)/2\rfloor} \sim p^d$ near threshold; the steep gradient at high $d$ provides strong selective pressure for distance improvements. The connectivity bonus $\alpha_{\mathrm{conn}} = 1.3$ for connected graphs rewards global graph structure associated with larger minimum-weight logical operators. The edge penalty $\beta = 0.5$ is calibrated so that, at the typical edge counts found in $n=15$ solutions ($|E| \approx 20$--$40$), the penalty term contributes roughly 10--15\% of the total reward, sufficient to prevent overcrowding without dominating the distance signal.

For \emph{hardware-adapted objectives}, the reduced exponent $d^{2.5}$ (rather than $d^3$) reflects that degree penalties already provide strong regularization; a softer distance gradient allows SA to explore a broader region of graph space before committing to high-degree configurations. The maximum-degree penalty coefficient~5 and average-degree coefficient~2 are set so that, for a typical solution with $\Delta(G) = 4$ and $\delta_{\mathrm{avg}} \approx 2.5$, the combined degree penalty is $5 \times 4 + 2 \times 2.5 = 25$ reward units, representing 15--20\% of a typical total reward.

For \emph{rate-distance}, \emph{cluster-state}, \emph{surface-like}, and \emph{connectivity} objectives, analogous design logic applies: penalty and bonus coefficients are set to keep each ancillary term at 10--20\% of the primary reward, ensuring that SA dynamics are driven primarily by the objective of interest.

\section{Fisher information per qubit across the converged code family}
\label{app:fisher-per-qubit}

\begin{figure}[t]
\centering
\includegraphics[width=\columnwidth]{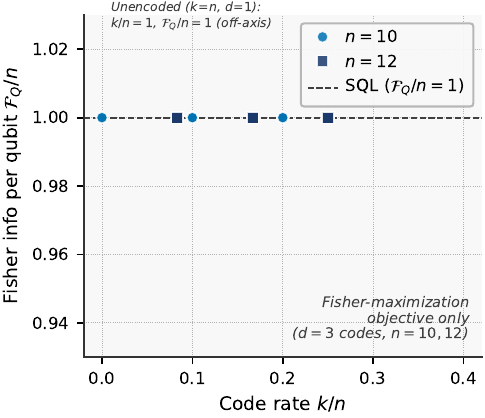}
\caption{Code parameter distributions ($\mathcal{F}_Q/n$ vs.\ rate $k/n$) for the Fisher Information objective across $n=10$ (circles) and $n=12$ (squares) qubit codes from 40 converged runs. All converged codes achieve $\mathcal{F}_Q/n \approx 1.0$ (SQL level) with distance $d=3$. The horizontal scatter in $k/n$ reflects the multiplicity of SQL-saturating equilibria reached by the search rather than a tradeoff between rate and metrological sensitivity, since the Fisher objective does not directly reward rate.}
\label{fig:fisher-per-qubit-appendix}
\end{figure}

The rate--Fisher tradeoff plane for the converged code family discussed in Sec.~III.B is summarized in \cref{fig:fisher-per-qubit-appendix}. The data collapse onto a single horizontal line at $\mathcal{F}_Q/n = 1$ demonstrates that every Nash equilibrium reached by the Fisher information objective saturates the standard quantum limit at the system sizes studied. The rate $k/n$ spans the interval $[0.08,\,0.25]$ across the 40 converged runs because the Fisher objective $f_{\mathrm{Fisher}}(G) = \mathcal{F}_Q(G)$ contains no explicit rate term; the realized rate at equilibrium is therefore determined by the connectivity of the discovered graph rather than by competition between objectives. This decoupling of metrological sensitivity and encoding rate is a structural feature of the single-objective Fisher game and is consistent with the interpretation of the discovered equilibria as a one-parameter family of SQL-saturating graph states.

\end{document}